\documentclass[12pt]{article}
\usepackage{authblk}
\usepackage[bookmarksnumbered, colorlinks, plainpages]{hyperref}
\usepackage{amsmath, amsthm, amscd, amsfonts, amssymb, graphicx, color, booktabs}
\usepackage[separate-uncertainty = true]{siunitx}
\usepackage[font=footnotesize,labelfont=bf]{caption}
\usepackage{fancyhdr}
\usepackage{cite}
\numberwithin{equation}{section}
\definecolor{email}{rgb}{0.00,0.00,0.84}
\begin{document}
\setcounter{page}{1}

\newcommand{\bra}[1]{\langle #1 \rvert}
\newcommand{\ket}[1]{\lvert #1 \rangle}

\pagestyle{fancy}
\renewcommand{\headrulewidth}{0pt}
\fancyhead{} 
\fancyhead[LE,LO]{P3H-24-016, TTP24-005}

\title{\large \bf 12th Workshop on the CKM Unitarity Triangle\\ Santiago de Compostela, 18-22 September 2023 \\ \vspace{0.3cm}
\LARGE NNLO QCD corrections to $\mathbf{\Delta \Gamma_s}$ in the $\mathbf{B_s-\overline{B}_s}$ system}

\author[1]{Marvin Gerlach}
\author[2]{Ulrich Nierste}
\author[3]{Pascal Reeck}
\author[4]{Vladyslav Shtabovenko}
\author[5]{Matthias Steinhauser}

\affil[1,2,3,5]{Karlsruhe Institute of Technology, Germany}
\affil[4]{University of Siegen, Germany}

\maketitle
\thispagestyle{fancy}
\begin{abstract}
This report summarises recent advances made in the calculation of the NNLO QCD corrections to the width difference $\Delta\Gamma_s$ in the $B_s-\overline{B}_s$ system. The inclusion of the effects due to current-current operators leads to an updated prediction of $\Delta\Gamma_s = \SI{0.076(17)}{\pico\second^{-1}}$, which narrows the gap between theory and experiment.
\end{abstract}

\section{Introduction}
The mixing of $B_{s}$ and $\overline{B}_{s}$ mesons is fully described by the off-diagonal elements of the self-energy matrix $\Sigma$ and a calculation of the corresponding matrix elements leads to theoretical predictions for the mass difference $\Delta M_s$ and the width difference $\Delta \Gamma_s$ of the mass eigenstates. The self-energy is related to the scattering matrix elements through
\begin{equation}
-i (2\pi)^4\delta^{(4)}(p_i-p_j) \Sigma_{ij} = \frac{1}{2 M_B} \bra{B_i} S \ket{B_j}.
\end{equation}
Within the Wigner-Weisskopf approximation, we can write down the Schrödinger equation for the two-state system as  \cite{Nierste:2009wg, Weisskopf:1930au, Lee:1957qq}
\begin{equation}
i \frac{\text{d}}{\text{d} t} \begin{pmatrix} \ket{B(t)}\\ \ket{\bar{B}(t)} \end{pmatrix} = \Sigma \begin{pmatrix}\ket{B(t)}\\ \ket{\bar{B}(t)}\end{pmatrix},
\end{equation}
where the Hermitian mass and decay width matrices $M$ and $\Gamma$ can be defined through $\Sigma = M - \frac{i}{2} \Gamma$. Diagonalising the matrix $\Sigma$ leads to the eigenstates, $B_H$ and $B_L$. The width difference between these states is given by
\begin{equation}
\begin{split}
\Delta \Gamma 
	&= - 2\lvert \Gamma_{12} \rvert \cos(\phi_{12}) + \mathcal{O}\left(\frac{\lvert\Gamma_{12}\rvert^2}{\lvert M_{12}\rvert^2}\right)\\
	&= \lvert \Sigma_{12} - \Sigma_{21}^* \rvert \cos(\phi_{12}) + \mathcal{O}\left(\frac{\lvert\Gamma_{12}\rvert^2}{\lvert M_{12}\rvert^2}\right),
\end{split}
\end{equation}
where the CP-violating phase $\phi_{12}$ is the phase difference between the phases of $M_{12}$ and $\Gamma_{12}$.

\section{Calculation overview}
The latest update from on the theoretical calculation of $\Delta\Gamma$ from Refs.~\cite{Gerlach:2021xtb, Gerlach:2022wgb,Gerlach:2022hoj} focused on reducing the perturbative uncertainties in the leading $\mathcal{O}((\Lambda_\text{QCD}/m_b)^0)$ terms. This is achieved through a matching calculation of a $\lvert\Delta B\rvert = 2$ matrix element calculated within effective $\lvert\Delta B\rvert = 1$ and $\lvert\Delta B\rvert = 2$ theories, where the high-energy and low-energy effects factorise into the matching coefficients and the operator matrix elements respectively. To obtain only the leading terms in $\Lambda_\text{QCD}/m_b$, the Heavy Quark Expansion (HQE) is used for the transition operator on the $\lvert\Delta B\rvert = 2$ side, which allows us to expand the operators in $\Lambda_\text{QCD}/m_b$ \cite{Khoze:1983yp, Shifman:1984wx, Khoze:1986fa, Chay:1990da, Bigi:1991ir, Bigi:1992su, Bigi:1993fe, Blok:1993va, Manohar:1993qn, Lenz:2014jha}. The matching calculation is done methodically by first calculating the imaginary part of the $B_s \rightarrow \overline{B}_s$ mixing amplitude in the two effective field theories, renormalising the results and then matching the coefficients of the $\lvert\Delta B\rvert = 2$ operators to the result from the $\lvert\Delta B\rvert = 1$ calculation.

For the calculation on the $\lvert\Delta B\rvert = 1$ side, we use the  Chetyrkin-Misiak-Münz (CMM) basis, which is particularly useful for automated calculations in our application as it circumvents all of the complications related to $\gamma_5$ in dimensional regularisation in our case. The Hamiltonian in the CMM basis \cite{Chetyrkin:1997gb} is given by
\begin{equation}
\begin{split}
\mathcal{H}^{\lvert\Delta B\rvert = 1} = \frac{4 G_F}{\sqrt{2}} \sum_{j=1}^2 C_j \left( V_{cb} V_{cs}^* P_j^{cc} + V_{cb} V_{us}^* P_j^{cu} + V_{ub} V_{cs}^* P_j^{uc} + V_{ub} V_{us}^* P_j^{uu} \right) \\
- \frac{4 G_F}{\sqrt{2}} V_{tb} V_{ts}^* \left( \sum_{j=3}^6 C_j P_j + C_8 P_8 \right)+ \sum C_{E_j} E_j + \text{h.c.},
\end{split}
\end{equation}
where $G_F$ is the Fermi constant and $V_{ij}$ are the CKM matrix elements. The operators $P_{1,2}$ are the current-current operators which couple to two up-type quarks as specified by the superscript. The penguin operators are $P_{3-6}$ and the operator $P_8$ is the chromomagnetic operator; all operators are defined in Ref.~\cite{Chetyrkin:1997gb}. The Wilson coefficients $C_i$ have been calculated to three-loop order in previous works \cite{Gambino:2003zm, Gorbahn:2004my, Gorbahn:2005sa}. Another issue related to dimensional regularisation with $d=4-2\epsilon$ dimensions is the appearance of so-called evanescent operators, which are of order $\epsilon$ and vanish if four-dimensional Dirac identities are applied. However, the evanescent operators mix with physical operators and need to be taken into consideration when renormalising bare amplitudes. Furthermore, if dimensional regularisation is used  for infrared divergences, the coefficients $C_{E_j}$ enter the calculation, see Sec.~\ref{sec:evan_op}.

To calculate the width difference $\Delta\Gamma$, the absorptive part of the scattering matrix element needs to be evaluated, which decomposes into a sum of terms with different CKM factors. This prompts us to decompose the $\lvert\Delta B\rvert = 2$ matching coefficients in an analogous fashion. From these considerations we can write the off-diagonal matrix element of the decay width in the $\lvert\Delta B\rvert = 1$ theory as
\begin{equation}
\Gamma_{12} = \frac{1}{M_B} \sum_{\alpha,\beta} \lambda_\alpha \lambda_\beta\, \text{Im}(\mathcal{M}_{\alpha\beta}),
\end{equation}
where $\lambda_\alpha \equiv V_{\alpha s}^* V_{\alpha b}$, and in the $\lvert\Delta B\rvert = 2$ theory as
\begin{equation}
\Gamma_{12} = - \frac{G_F^2 m_b^2}{24 \pi^2 M_B} \sum_{\alpha,\beta} \lambda_\alpha \lambda_\beta \left[H^{\alpha\beta}  \bra{B}Q \ket{\bar{B}} + \tilde{H}_S^{\alpha\beta}  \bra{B} \tilde{Q}_S \ket{\bar{B}} \right] + \mathcal{O}\left(\frac{\Lambda_\text{QCD}}{m_b}\right),\label{eq:dGamma_full}
\end{equation}
where $M_B$ is the mass of the B meson. In the context of the HQE, the matching coefficients $H^{\alpha\beta}$ and $\tilde{H}_S^{\alpha\beta}$ are calculated as expansions in $z \equiv m_c^2/m_b^2$. The physical operators of the $\lvert \Delta B \rvert = 2$ transition operator are given by
\begin{align}
Q &= \left(\bar{b}_i \gamma^\mu \,(1-\gamma_5)\, s_i\right) \left(\bar{b}_j \gamma_\mu \,(1-\gamma_5)\, s_j\right), \\
\tilde{Q}_S &= \left(\bar{b}_i \,(1+\gamma_5)\, s_j\right) \left(\bar{b}_j \,(1+\gamma_5)\, s_i\right).
\end{align}
As alluded to previously, the low-energy and high-energy physics factorise with the matching coefficients $H^{\alpha\beta}$ and $\tilde{H}_S^{\alpha\beta}$ containing the perturbative high-energy physics that is the main goal of the theoretical calculation described here. The low-energy behaviour captured in the operator matrix elements of the physical operators needs to be extracted from either QCD sum rules \cite{Ovchinnikov:1988zw, Reinders:1988aa, Korner:2003zk, Mannel:2011iqd, Grozin:2016uqy, Kirk:2017juj, King:2019lal, King:2021jsq} or lattice QCD calculations \cite{Davies:2019gnp,Dowdall:2019bea} and is used as an input in the prediction of $\Delta \Gamma$.

\section{Dimensional regularisation and evanescent operators}\label{sec:evan_op}
In dimensional regularisation we regularise ultraviolet (UV) poles by choosing the dimension to be $d=4-2\epsilon$ where $\epsilon$ is a small parameter that is set to zero in the renormalised amplitudes. Evanescent operators are operators which are of order $\epsilon$ and vanish in four dimensions due to four-dimensional identities of the Dirac algebra, e.g.~Chisholm identities as well as Fierz identities. However, their Wilson coefficients mix with the physical operators and consequently they become important in the renormalisation procedure.

An additional complication arises when infrared (IR) poles are also regularised with $\epsilon=\epsilon_\text{UV}=\epsilon_\text{IR}$, which means that IR poles and finite terms from evanescent operators remain after renormalisation and only cancel in the matching between the $\lvert\Delta B \rvert = 1$ and $\lvert\Delta B \rvert = 2$ sides. Moreover, lower orders in $\alpha_s$ need to be calculated to higher orders in $\epsilon$ to extract all required matching coefficients because the matching equation contains IR poles. To obtain NNLO matching coefficients, we need to match up to $\mathcal{O}(\epsilon^1)$ for NLO and up to $\mathcal{O}(\epsilon^2)$ for LO.

The $\lvert\Delta B\rvert = 2$ basis contains a peculiar operator which arises from a linear combination of physical operators and whose evanescent piece needs to be mentioned. In addition to the vector and pseudoscalar operators $Q$ and $\tilde{Q}_S$ defined above, we have the corresponding operators with different colour structures,
\begin{align}
\tilde{Q} &= \left(\bar{b}_i \gamma^\mu \,(1-\gamma_5)\, s_j\right) \left(\bar{b}_j \gamma_\mu \,(1-\gamma_5)\, s_i\right), \\
Q_S &= \left(\bar{b}_i \,(1+\gamma_5)\, s_i\right) \left(\bar{b}_j \,(1+\gamma_5)\, s_j\right).
\end{align}
The operators $Q$ and $\tilde{Q}$ are equal in four dimensions due to a Fierz identity; thus their difference is evanescent, i.e.~of order $\epsilon$. However, there is another linear relation between the physical operators, which leads to
\begin{equation}
R_0 = \frac12 Q + Q_S + \tilde{Q}_S,
\end{equation}
an operator whose matrix element $\langle R_0 \rangle^{(0)}$ is $\Lambda_\text{QCD}/m_b$ suppressed in our process \cite{Beneke:1996gn}. However, this operator also has an evanescent part which is unsuppressed. Hence, the operator needs to be properly renormalised and in particular its $\epsilon$-finite renormalisation constants, which are needed to remove any dependence of the physical amplitude on the evanescent parts, have to be implemented.

\section{Technical details on the calculation}
The kinematics of the calculation are such that the external quarks are on-shell with $p_b^2 = m_b^2$ for the bottom quark while $p_s = 0$ is chosen for the massless strange quark. Internal up and down quarks are also taken as massless while the charm quark is given a mass $m_c$. Diagrams for the calculation are generated using \texttt{QGRAF} \cite{Nogueira:1991ex}. For the insertion of Feynman rules and identification of topologies, \texttt{tapir} \cite{Gerlach:2022qnc} and \texttt{exp} \cite{Harlander:1998cmq,Seidensticker:1999bb} are employed and the integrals are reduced with the integration by parts technique using \texttt{FIRE} \cite{Smirnov:2019qkx} or \texttt{Kira} \cite{Maierhofer:2017gsa, Klappert:2020nbg}. Since only the imaginary part of the integrals is relevant to the calculation of $\Delta \Gamma$, only those master integrals which have a physical cut will contribute, i.e.~all masters where all cuts go through a bottom quark line can be discarded. The resulting master integrals can finally be evaluated using \texttt{HyperInt} \cite{Panzer:2014caa}. 

The update on the theoretical calculation in Refs.~\cite{Gerlach:2021xtb, Gerlach:2022wgb, Gerlach:2022hoj} makes use of the fact that the numerical results converge quickly in an expansion in $z\equiv m_c^2/m_b^2$ and the NNLO QCD corrections are evaluated to $\mathcal{O}(z)$ by expanding naively in $m_c$. This is only valid for the diagrams which do not contain a closed charm loop, but those contributions have already been calculated in previous works \cite{Asatrian:2017qaz, Asatrian:2020zxa}. Combining these results, the contributions to $\Delta \Gamma$ have been updated with the operators and loop orders as shown in Tab.~\ref{tab:dGamma_updated_ops}.
\begin{table}
\centering
\begin{tabular}{@{}  l  l  l @{}}
\toprule
Contribution & Previous results & Refs.\cite{Gerlach:2021xtb, Gerlach:2022wgb, Gerlach:2022hoj} \\ \midrule
$P_{1,2} \times P_{3-6}$ & 2 loops, $z$-exact, $n_f$-part only \cite{Asatrian:2017qaz, Asatrian:2020zxa}  & 2 loops, $\mathcal{O}(z)$, full\\
$P_{1,2} \times P_8$ & 2 loops, $z$-exact, $n_f$-part only \cite{Asatrian:2017qaz, Asatrian:2020zxa}&  2 loops, $\mathcal{O}(z)$, full\\
$P_{3-6} \times P_{3-6}$ & 1 loop, $z$-exact, full \cite{Beneke:1996gn} &  2 loops, $\mathcal{O}(z)$, full\\
$P_{3-6} \times P_{8}$ & 1 loop, $z$-exact, $n_f$-part only \cite{Asatrian:2017qaz, Asatrian:2020zxa}  &  2 loops, $\mathcal{O}(z)$, full\\
$P_{8} \times P_{8}$ & 1 loop, $z$-exact, $n_f$-part only \cite{Asatrian:2017qaz, Asatrian:2020zxa} &  2 loops, $\mathcal{O}(z)$, full\\
$P_{1,2} \times P_{1,2}$ & 3 loops, $\mathcal{O}(\sqrt{z})$, $n_f$-part only \cite{Asatrian:2017qaz, Asatrian:2020zxa} &  3 loops, $\mathcal{O}(z)$, full\\
\bottomrule
\end{tabular}
\caption{Updated contributions to the theoretical value of $\Delta \Gamma$. ``Full'' contributions refers to the fact that both the fermionic and non-fermionic pieces have been calculated.}\label{tab:dGamma_updated_ops}
\end{table}

\section{Results}
With the results in Refs.\cite{Gerlach:2021xtb, Gerlach:2022wgb, Gerlach:2022hoj}, $\Delta\Gamma$ has now been updated to include the NNLO QCD corrections stemming from diagrams with two insertions of the current-current operators on the $\lvert\Delta B\rvert = 1$ side. Contributions from penguin operators and the chromomagnetic operator are also updated to higher accuracy than in previous calculations. One detail that has a large impact on the numerical result and in particular the renormalisation scale dependence is the choice of the mass scheme. Due to the renormalon ambiguity in the on-shell mass definition \cite{Beneke:1998ui}, a better convergence behaviour is achieved through converting the on-shell, i.e.~pole mass ratio $z=(m_c^\text{pole}/m_b^\text{pole})^2$ in the matching coefficients to the corresponding ratio in the $\overline{\text{MS}}$ scheme, $\overline{z}\equiv(\overline{m}_c/\overline{m}_b)^2$. In addition to this, there is also another factor of $m_b^2$ in Eq.~\eqref{eq:dGamma_full} multiplying $\Delta\Gamma$. In a first step one usually employs a pole mass, but subsequently trades it for an $\overline{\text{MS}}$   or the potential-subtracted (PS) mass \cite{Beneke:1998rk}, which have better infrared properties. All three choices have been considered to estimate the renormalisation scale dependence of the NNLO result.

To further improve the numerical accuracy of the result, it is helpful to consider the ratio $\Delta\Gamma/\Delta M$. In this ratio the dependence on $\lvert V_{cb}\rvert$ drops out and most of the dependence on the bag parameters is also removed. Since $\Delta M$ is already known to NNLO in QCD \cite{Buras:1990fn}, $\Delta\Gamma/\Delta M$ can also be calculated to NNLO. The results for the phenomenologically interesting ratio in the different mass schemes of the overall $m_b^2$ factor are given by
\begin{align}
\left.\frac{\Delta\Gamma_s}{\Delta M_s}\right\rvert_{\text{pole}} &= \left (3.79 \, \substack{+0.53 \\ -0.58}_\text{scale} \, \substack{+0.09 \\ -0.19}_{\text{scale}, 1/m_b} \pm 0.11_{B\tilde{B}_S} \pm 0.78_{1/m_b} \pm 0.05_\text{input} \right) \times 10^{-3},\\
\left.\frac{\Delta\Gamma_s}{\Delta M_s}\right\rvert_{\overline{\text{MS}}} &= \left ( 4.33 \, \substack{+0.23 \\ -0.44}_\text{scale} \, \substack{+0.09 \\ -0.19}_{\text{scale}, 1/m_b} \pm 0.12_{B\tilde{B}_S} \pm 0.78_{1/m_b} \pm 0.05_\text{input} \right) \times 10^{-3},\\
\left.\frac{\Delta\Gamma_s}{\Delta M_s}\right\rvert_{\text{PS}} &=  \left ( 4.20 \, \substack{+0.36 \\ -0.39}_\text{scale} \, \substack{+0.09 \\ -0.19}_{\text{scale}, 1/m_b} \pm 0.12_{B\tilde{B}_S} \pm 0.78_{1/m_b} \pm 0.05_\text{input} \right) \times 10^{-3},
\end{align}
where the subscripts on the uncertainties indicate their origin \cite{Gerlach:2022wgb}. The scale dependence refers to the remaining dependence on the renormalisation scale for the leading and sub-leading terms in $\Lambda_\text{QCD}/m_b$ respectively. The biggest contribution labelled with $1/m_b$ stems from the uncertainty on the hadronic matrix elements of $\Lambda_\text{QCD}/m_b$ operators. The uncertainties of the bag parameters $B$ and $\tilde{B}_S$ also give a significant contribution and all other uncertainties of numerical input parameters are included in the input uncertainty.

\begin{figure} [htb!]
 \centering
 \includegraphics[width=0.75\textwidth]{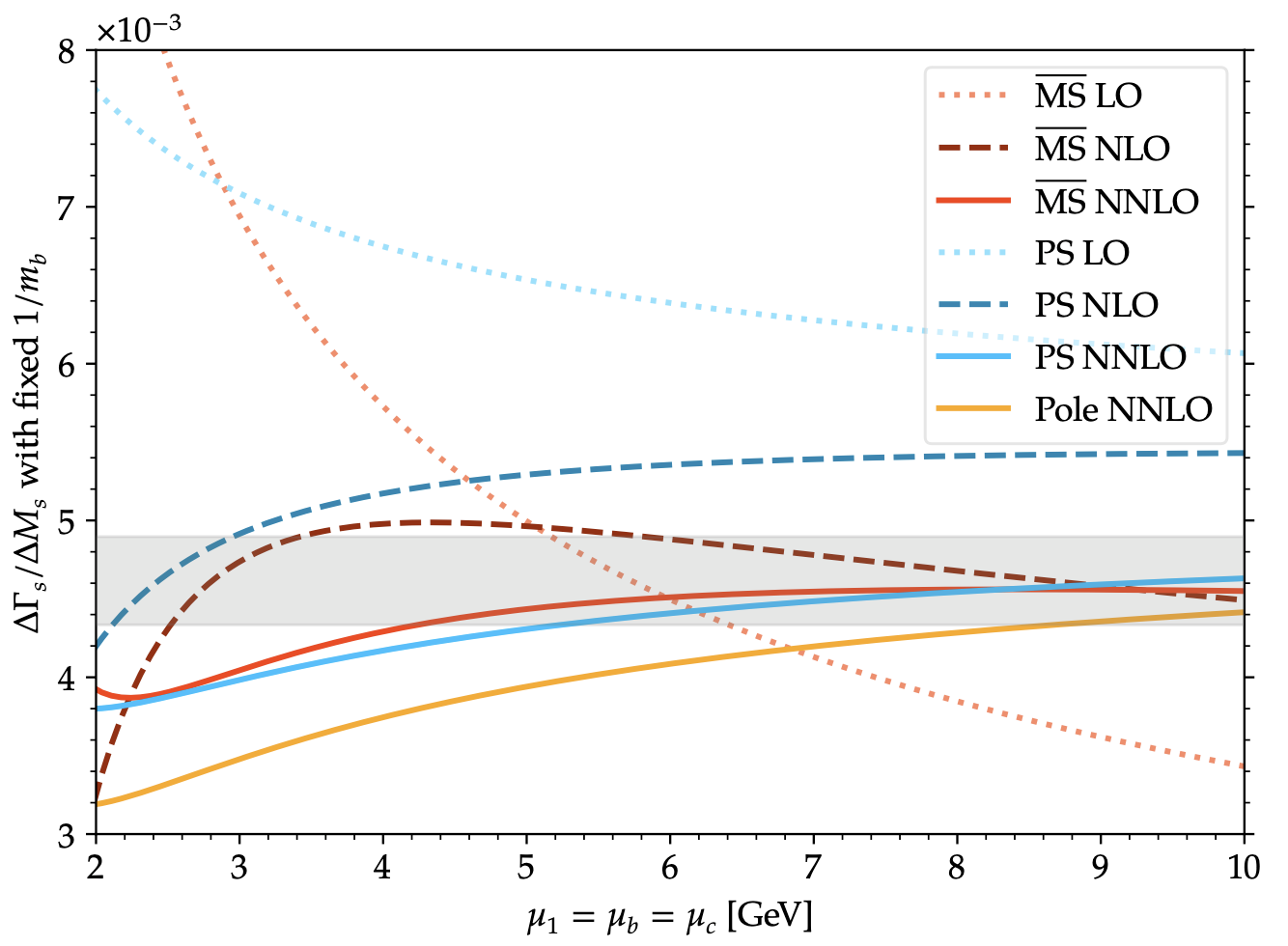}
 \caption{The renormalisation scale dependence of $\Delta\Gamma/\Delta M$ in the $B$-$\overline{B}$ system as calculated in Ref.~\cite{Gerlach:2022wgb}. Note that the renormalisation scale of the matching calculation, $\mu_1$, is varied simultaneously with the renormalisation scales of the $\overline{\text{MS}}$ bottom and charm masses, $\mu_b$ and $\mu_c$ respectively. Since the focus is on the leading terms in $\Lambda_\text{QCD}/m_b$, only the renormalisation scale of those terms is varied while the scale of the $1/m_b$ terms is kept fixed.}
 \label{fig:dGdM_scale}
 \end{figure}

The renormalisation scale dependence from which the scale errors of the leading term in $\Lambda_\text{QCD}/m_b$ are determined is shown in Fig.~\ref{fig:dGdM_scale}. It is reassuring to observe that the renormalisation scale dependence is indeed improved by the inclusion of the NNLO corrections stemming from current-current operators. Moreover, it is clear from the plot that the pole scheme leads to inaccurate results due to its large deviation from the other two schemes. This feature is commonly observed and conventionally ascribed to the renormalon problem of the pole mass \cite{Beneke:1994sw, Bigi:1994em}.
 
 Using the experimental value for $\Delta M_s$ \cite{LHCb:2021moh},
 \begin{equation}
\Delta M_s^\text{exp} =  \SI{17.7656(57)}{\pico\second^{-1}},
\end{equation}
the theoretical prediction for $\Delta \Gamma_s$ is updated to be
\begin{equation}
\Delta \Gamma_s^\text{th} = \SI{0.076(17)}{\pico\second^{-1}}.
\end{equation}
Note that only the $\overline{\text{MS}}$ and PS results were used to obtain this final number. Comparing this result to the experimental value \cite{HFLAV:2022esi},
\begin{equation}
\Delta \Gamma_s^\text{exp}  \SI{0.084(5)}{\pico\second^{-1}},
\end{equation}
we conclude that the theoretical uncertainty is about three times as large as the experimental one.

\section{Conclusion}
The first step towards a NNLO calculation of the QCD corrections to $\Delta\Gamma_s$ has now been completed. The theoretical predictions agree well with the experimental measurements within the respective uncertainties. Calculations to further reduce the theoretical uncertainties are already underway and are aiming to improve the accuracy by including higher-order terms in $z$ as well as the penguin operator contributions at NNLO.


\section*{Acknowledgments}
The author would like to thank Matthias Steinhauser, Ulrich Nierste and Vladyslav Shtabovenko for their support and collaboration. This research was supported by the Deutsche Forschungsgemeinschaft (DFG, German Research Foundation) under grant 396021762 -- TRR 257 ``Particle Physics Phenomenology after the Higgs Discovery''.

\bibliographystyle{JHEP}
\bibliography{references}

\end{document}